\documentclass[prl,twocolumn,showpacs,superscriptaddress]{revtex4}
\usepackage[english]{babel}
\usepackage{graphicx}
\usepackage{xspace}
\usepackage{amsmath,amssymb}

\begin{document}

\bibliographystyle{prsty}

\title {Comment on "Raman spectroscopy study of Na$_x$CoO$_2$
and superconducting Na$_x$CoO$_2\cdot y$H$_2$O"}

\author{P.~Lemmens}
\email{ p.lemmens@tu-bs.de}\affiliation{Institute for Physics of Condensed Matter, TU Braunschweig, D-38106
Braunschweig, Germany}
\author{P.~Scheib}
\affiliation{Institute for Physics of Condensed Matter, TU Braunschweig, D-38106
Braunschweig, Germany}
\author{Y.~Krockenberger}
\affiliation{Max-Planck-Institute for
Solid State Research, Heisenbergstr.~1, 70569 Stuttgart, Germany}
\affiliation{Institute of Materials Science, TU Darmstadt,
Petersenstr.~23, D-64287 Darmstadt, Germany}
\author{L.~Alff}
\affiliation{Institute of Materials Science, TU Darmstadt,
Petersenstr.~23, D-64287 Darmstadt, Germany}
\author{F.~C.~Chou}
\affiliation{Center for Materials Science and Engineering, MIT, Cambridge, MA 02139, USA}
\author{C.~T.~Lin}
\affiliation{Max-Planck-Institute for
Solid State Research, Heisenbergstr.~1, 70569 Stuttgart, Germany}
\author{H.-U. Habermeier}
\affiliation{Max-Planck-Institute for
Solid State Research, Heisenbergstr.~1, 70569 Stuttgart, Germany}
\author{B. Keimer}
\affiliation{Max-Planck-Institute for
Solid State Research, Heisenbergstr.~1, 70569 Stuttgart, Germany}

\begin{abstract}

The effect of surface degradation of the thermolectric cobaltite on Raman spectra is
discussed and compared to experimental results from Co$_3$O$_4$ single crystals. We
conclude that on NaCl flux grown Na$_x$CoO$_2$ crystals a surface layer of Co$_3$O$_4$
easily forms that leads to the observation of an intense phonon around 700~cm$^{-1}$
[Phys. Rev. B \textbf{70}, 052502 (2004)]. Raman spectra on freshly cleaved crystals
from optical floating zone ovens do not show such effects and have a high frequency
phonon cut-off at approximately 600~cm$^{-1}$ [Phys. Rev. Lett \textbf{96}, 167204
(2006)]. We discuss the relation of structural dimensionality, electronic correlations
and the high frequency phonon cut-off of the thermolectric cobaltite.

\end{abstract}
\date{\today}
 \maketitle


Raman scattering is a well established probe for structural and electronic properties of
solids as, e.g. compositional and symmetry information can be gained from the number and
frequency of the observed phonon modes \cite{cardona}. On the other side its high
surface sensitivity may also lead to challenges in sample preparation. The cobaltite
Na$_x$CoO$_2\cdot y$H$_2$O is a correlated electron system with an enormous thermopower
for large $x>0.7$ and superconductivity for smaller $x=1/3$ and hydration, $y=1.3$. Due
to the large mobility of Na on different sites and the mixed nominal Co valency
Na$_x$CoO$_2$ has a complex defect chemistry. In the presence of CO$_2$ and humidity
surface layers are formed that consists of, e.g. CoCO$_3$, Na$_2$CO$_3$ and Co$_3$O$_4$.
The latter compound is also used as an ingot material in sample preparation
\cite{highT}.


\begin{figure}[b]
     \centering
     \includegraphics[height=4.5cm]{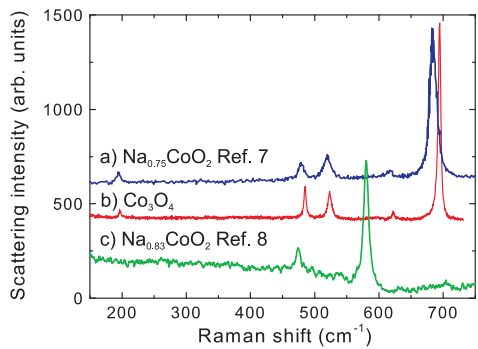}
          \caption{Raman scattering spectra of (a) NaCl flux grown Na$_{0.75}$CoO$_2$ at RT
          (Ref.~\onlinecite{shi}), (b) Co$_3$O$_4$ at $T=200$\,K and (c) TFZ grown Na$_{0.83}$CoO$_2$
           at $T=90$\,K (Ref.~\onlinecite{lemmens}).}
 \end{figure}

The preparation of large single crystals has been reported from optical traveling
floating zone (TFZ) ovens \cite{tfz} and from NaCl flux (NaCl, Na$_2$CO$_3$, and
B$_2$O$_3$ in varying compositions) \cite{flux}. TFZ grown crystals can easily be
cleaved, while samples from NaCl flux are washed-out from the flux in water. The latter
step may lead to a Na nonstoichiometry. Evidence for degradation and crystallographic
changes of Na$_x$CoO$_2$ and Na$_x$CoO$_2\cdot y$H$_2$O on time scales from minutes to
weeks exist in literature. \cite{iliev,fisher}.


In a recent Raman scattering investigation of NaCl-flux grown Na$_x$CoO$_2$ crystals,
Shi {\em et al.}~have reported Raman spectra that show 5 phonons with in-plane
polarization \cite{shi}, see Fig.~1, curve a). These modes are attributed to five Raman
active modes corresponding to displacements of sodium and oxygen \cite{iliev}. In
contrast to these data Raman scattering investigations on freshly cleaved TFZ grown
crystals give only two modes with larger intensity \cite{lemmens}. These modes are
attributed to oxygen in-plane and out-of-plane displacements. While the non-observance
of the low frequency Na modes is attributed to disorder on the partially occupied Na
sites \cite{huang}, the vibrations of oxygen within the CoO$_{2}$ layers should have
characteristic frequencies. Indeed a linear frequency shift of the highest frequency,
out-of-plane mode at 590\,cm$^{-1}$ by 5\% has been found with increasing Na content in
the TFZ crystals \cite{lemmens}. The shift implies that the oxygen modes only weakly
depend on the stacking of the CoO$_2$ layers and the occupation of Na sites that
characterize the ($\alpha, \beta, \gamma$ type) crystal structure \cite{huang}. With
this respect the compound can be considered as two-dimensional and the evolution of
electronic correlations with doping dominates the phonon frequency \cite{sherman}. Our
experiments are further supported by recent inelastic X-ray scattering that show a bend
over of the highest phonon branch at about 70\,meV $\equiv$ 583\,cm$^{-1}$ \cite{rueff}.

In contrast, the three-dimensional Co$_3$O$_4$ has a very intense Raman mode at a higher
frequency (690\,cm$^{-1}$), i.e. in the same frequency regime as Raman data \cite{shi}
of NaCl-flux grown Na$_x$CoO$_2$ crystals. In Fig.~1 we show respective spectra. The
small frequency shift and broadening of curve a) compared to b) is attributed to an
oxygen deficiency or a small thickness of the surface layer. Similar data on Co$_3$O$_4$
have been reported earlier by Hadjiev {\em et al.} \cite{hadjiev} and more recently by
Qu {\em et al.} discussing phase separation \cite{qu}. We conclude that the Raman data
\cite{shi} of NaCl-flux grown Na$_x$CoO$_2$ are most probably interfered by a
degradation of the sample leading to a surface layer of Co$_3$O$_4$. We highlight that
although from symmetry analysis the same number of Raman active modes are expected, the
frequency of the modes in Co$_3$O$_4$ and Na$_x$CoO$_2$ differ considerably. The intense
Co$_3$O$_4$ mode at 690\,cm$^{-1}$ can be used as a
quality measure of cobaltates in thermoelectric applications.\\

\textbf{Acknowledgement: }We acknowledge support by the DFG within the project
Le~967/4-1 and the ESF program \emph{Highly Frustrated Magnetism}.



\begin{thebibliography}{99}
\bibitem{highT} At high temperatures ($T>$850$^\circ$C) Na$_2$O has a large volatility.
As a result the more stable Co$_3$O$_4$ forms instead of Na$_x$CoO$_2$ with smaller or
varying $x$. At ambient conditions the remaining Na$_2$O together with CO$_2$ and H$_2$O
forms Na$_2$CO$_3$.
\end{thebibliography}
\end{document}